\documentclass[12pt]{article}

\usepackage[round]{natbib}
\usepackage{amsmath, amssymb} % AMS packages
\usepackage{color}
\usepackage{algorithm}
\usepackage{algorithmic}
\usepackage{graphicx}

\usepackage[normalem]{ulem}
\usepackage{xcolor}
\definecolor{gray}{gray}{0.5}

\begin{document}

\title{Bayesian model selection for the glacial-interglacial cycle}

\author{Jake Carson$^1$\footnote{Email: Jake.Carson@warwick.ac.uk}, Michel Crucifix$^2$,\\ Simon Preston$^3$, Richard Wilkinson$^4$}

\date{ }

\maketitle

\begin{center}
{\small
 $^1$ University of Warwick, Department of Statistics, UK. \\
 $^2$ Universit\'e catholique de Louvain, Earth and Life Institute, Belgium. \\
 $^3$ University of Nottingham, School of Mathematical Sciences, UK. \\
 $^4$ University of Sheffield, School of Mathematics and Statistics, UK. \\}
\end{center}

\begin{abstract}

A prevailing viewpoint in palaeoclimate science is that a single palaeoclimate record contains insufficient information to discriminate between most competing explanatory models.
Results we present here suggest the contrary.  
%Recent developments in Monte
%Carlo methodology, combined with advances in computer power, mean that for a wide class of
%phenomenological models, we are now able to perform simultaneous filtering, calibration,
%and Bayesian model selection.  [Filtering and calibration here are tangential to the main
%message - shall we cut mention of these here?]
Using SMC$^2$ combined with novel Brownian bridge type proposals for the state trajectories, we show that even with relatively short time series it is possible to estimate Bayes factors to sufficient accuracy to be able to select between competing models. 
The results show that Monte Carlo methodology and
computer power have now advanced to the point where a  full Bayesian analysis for a 
wide class of conceptual climate models is now possible.
The results also highlight a problem with estimating the chronology of
the climate record prior to further statistical analysis, a practice
which is common in palaeoclimate science.  Using two datasets based on
the same record but with different estimated chronologies results in
conflicting conclusions about the importance of the orbital forcing on
the glacial cycle, and about the
internal dynamics generating the glacial cycle, even though the
difference between the two estimated chronologies is consistent with
dating uncertainty.  This highlights a need for chronology estimation
and other inferential questions to be addressed in a joint statistical
procedure.

\end{abstract}


\section{Introduction}
\label{S:Intro}

Throughout the Pleistocene the Earth's climate has fluctuated between cold periods, in which glaciers expanded, and warm
periods in which the glaciers retreated \citep{Shackleton84ab}. 
The prevailing theory is that these  glacial--interglacial (GIG) cycles are driven, or at least partly
controlled, by changes in the Earth's orbit around the Sun which  affects the distribution of the incoming solar
radiation (or ``insolation''). The Earth's orbit undergoes cyclic changes in its eccentricity (degree of deviation from circular), obliquity (angle between the Earth's axis and the orbital plane), and its precession (which determines when the Earth is closest to the Sun, thus controlling the length of the seasons), and the value of each determines how much energy the Earth receives from the Sun. Although it is the combination of these signals that controls the incoming energy, many studies have examined the period and magnitude of each mode of variation to see if any one of these  can be considered to be the primary driver of the glacial-interglacial cycle.  
Opinion about which is the key aspect forcing the GIG cycle is varied and contradictory, with, for example, \citet{Huybers2005} arguing for obliquity,  \citet{Lisiecki2010} for eccentricity, and \citet{Huybers2011} for a combination of precession and obliquity.
These studies all analysed  features of the insolation signal using significance tests to assess 
whether the phases of each orbital parameter are highly
correlated with estimates of the glacial ``termination times'' (marking where individual glacial cycles finish), or  whether the
insolation signal was anomalously large at termination times.
Differences in the details about how these
 tests are constructed appear to substantially affect the
conclusions, with different studies finding different orbital characteristics being
of primary importance.

In contrast, it has long been  recognised that despite  the orbital control, the glacial-interglacial cycle is not entirely  predictable \citep{Imbrie80, Raymo97}. 
It has been shown that this lack of predictability may  emerge from the interactions between the orbital forcing and internal dynamics \citep{Crucifix2013}. To study these interactions,  Earth's climate must be modelled mathematically as a dynamical system, forced by the variation in the insolation. Atmospheric variability can be represented by stochastic processes, resulting in  stochastic differential equation (SDE) models.  The challenge is to  identify both what the appropriate forcing of the system is, and which is the best  mathematical representation of the climate's internal dynamics. 
Because investigating the dynamical effects associated with the emergence and stability of glacial cycles requires simulation over very long time-scales, it rules out the use of many complex Earth system models which  simulate the numerous physical mechanisms that affect the Earth's climate response. Instead, effort has focused on using  simpler
conceptual/phenomenological models, which typically involve just a few differential equations representing hypothesised relationships between parts of the climate.

The task of choosing between models is complicated  by the nature of the data available. There are no reliable direct measurements of either the Earth's climate or of the extent of the glaciers before the 19th century. Climate proxy records, such as the ratio of oxygen isotopes  $^{18}$O and $^{16}$O (referred to as $\delta ^{18}$O) measured in the calcite shells of foraminifera  preserved in temporally stratified layers on the ocean floor, are instead used to construct estimates of climate and ice extent \citep{emiliani55, shackleton67}. These data are noisy, and contain uncertainties on both the measured climate proxy, and on the date relating to that measurement.
Given this noise level,  models representing very different properties or
bifurcation structures may all be found to fit the data reasonably well when judged by eye (see \citet{Crucifix2013} for a recent account).  
As a consequence, a common viewpoint  is that the information contained in a single proxy record is not sufficient  to distinguish between the numerous proposed models (see, e.g., \citet{Roe1999}).

In this paper we develop a fully Bayesian approach that simultaneously estimates model parameters, the relative contribution of each aspect of the orbital forcing, and chooses between models by estimating Bayes factors. 
The statistical difficulty  in making inferences
from partially and noisily observed trajectories of forced non-linear SDEs lies in the computation of various posterior quantities. 
Inference for SDEs is particularly challenging because the transition density,
and therefore the likelihood function, is intractable (meaning it is not available in
closed form).
A powerful tool for time-structured problems with intractable likelihoods is the particle
filter, and in this paper we employ the SMC$^2$ approach recently introduced by
\citet{Chopin2012}. This is a pseudo-marginal algorithm that embeds a particle filter
within a sequential Monte Carlo algorithm to do joint state and parameter estimation. A
major advantage of SMC$^2$ over  competing methods, such as particle MCMC
\citep{Andrieu2010}, is that it allows for easy estimation of the model evidence, which we
exploit to provide estimates of the Bayes factors. 

A naive implementation of SMC$^2$ fails due to extreme particle degeneracy, but we show that by introducing novel guided Brownian bridge type proposals,   particle diversity with more evenly distributed weights can be maintained. When efficiently parallelised on a GPU, this allows inference to be performed in reasonable time (3-4 days on a single core, or 3-4 hours on a Tesla K20 GPU for the results in Section \ref{S:Results}). A surprising result is that even though the Monte Carlo error in our Bayes factor estimates is considerable, the Bayes factors are sufficiently large, even for short time series of data, that we can still distinguish between the competing models. The computational difficulty of estimating Bayes factors for models of this complexity pushes existing Monte Carlo methods and computer power towards the limit of what is possible.

Previous authors have also attempted model selection experiments for the GIG cycle, but with various limitations compared to our approach. 
\citet{Roe1999} compared deterministic models plus autoregressive process noise using an F-test and found no support for any one model over any other. 
\citet{Feng2015} used Bayesian model selection to select between
competing forcing functions over the Pleistocene, concluding that obliquity
influences the termination times over the entire Pleistocene, and that precession also has
explanatory power following the mid-Pleistocene transition. Their approach  requires knowledge of the likelihood function, which heavily restricts
the class of models that can be compared, in particular, ruling out the use of SDE models.   As in the previously mentioned hypothesis tests, they also begin by discarding most of the data and using a  summary consisting of  just the termination times  ($\sim 12$ over the past 1 Myr), which  is necessary as the low-order deterministic models used do not fit well to the complete dataset. They also only sample parameter values from the prior, leading to poor numerical efficiency. 
Finally, \citet{Kwasniok2013} compares conceptual models over the last glacial period using the Bayesian information criterion. The likelihood of each model is estimated using an unscented Kalman filter (UKF) \citep{Wan2000}. Whilst this approach focussed on a smaller time horizon than our application, it can be applied using the data and models in this paper. However, the Gaussian approximation used by the UKF, whilst working well for filtering, is unproven for  parameter estimation and  model selection, and the particle filter  offers a more natural approach for non-linear dynamical systems.

In contrast, our approach makes full use of the data not just the termination times, characterises  parametric uncertainty rather than plugging-in estimates, and quantifies the evidence in favour of each model through Bayes factor estimates with no approximation other than that incurred by the  Monte Carlo approximation. 
We will show that it is possible to  jointly estimate the state trajectory (a three dimensional vector over 800 time points), model parameters (up to 16 in one of the models), and estimate marginal likelihoods (allowing calculation of Bayes factors)
even using relatively
short time series of data, 
and that there is enough information in the data to choose between candidate conceptual models, including assessing the importance of the various orbital characteristics to the
glacial--interglacial cycle.
 
The paper is structured as follows. In Section \ref{Sec:Data}, we describe  the
data used, and the
models of the astronomical forcing,  the Earth's climate 
dynamics, and the proxy observations.  Section
\ref{S:Meth} contains a description of the Bayesian approach, a brief review of the
particle-filter, and we introduce our approach to inference, describing in detail how to avoid particle degeneracy using Brownian bridge type proposals.  In Section \ref{S:Results} we present 
a simulation study to assess the performance of the algorithms on synthetic data, and an  
analysis of a $\delta^{18}$O dataset.  In Section \ref{S:Conclusion} we offer
some thoughts on the practical implementation of the particle filter methods for such
problems, discuss the scientific conclusions, and suggest some future directions for
research.

\section{Data and Models}
\label{Sec:Data}

Our approach to understanding the dynamical behaviour of the palaeoclimate involves four
components: data consisting of palaeoclimate records;  models of the climate;  drivers of
the climate (such as CO$_2$ emissions or, more pertinently for palaeoclimate, the orbital forcing); and  a statistical model relating these three  components.
 In this paper we develop  the statistical methodology necessary for
combining these components, which we hope will allow palaeoclimate
scientists to study hypotheses in a statistically rigorous way. That is to say, given some data and a selection of models, we
show how to fit these models, and to assess which model is best supported by the data.
Scientific aspects of the approach can, and we hope will, be improved upon by using
different datasets and richer models.

\subsection{Data}
\label{SS:Data}

The ratio between the oxygen isotopes $^{18}$O and $^{16}$O, known as $\delta^{18}$O, 
reflects a combination of effects associated with changes in ocean temperature and sea-level 
\citep{emiliani55, shackleton67}. 
Broadly speaking, larger values of $\delta^{18}$O 
indicate a colder climate with greater ice volume.
Time series of measured  $\delta^{18}$O recorded in the calcite shells of foraminifera, are used as a proxy record for temperature and ice-extent, and provide a picture of the Earth's recent glaciations, particularly when combined with other information. %The first indices of Northern Hemisphere glaciation since 
The data we use are  measurements of $\delta^{18}$O from different depths in sediment cores extracted as part of the  Ocean Drilling Programme (ODP).
In climatology, a set of such measurements is known as a ``record'', and an average
over multiple records  is known as a ``stack'' \citep{imbrie84}.  The $\delta^{18}$O in deeper parts of a
core correspond to climate conditions further back in time.  Beyond monotonicity,
there is unfortunately no simple relationship between core depth and age.  This is because the
accumulation of sediment  results from a combination of 
complicated physical processes,
including sedimentation (which occurs at variable rates), erosion, and core
compaction.  A model for the relationship between depth and age is known as an ``age
model''.
 A common strategy in developing an age
model is to align features of records to important events visible in the core, such as magnetic reversals,
whose dates are accurately known from other sources \citep{Shackleton1990}.
In \citet{Huybers2007}, for instance, ``age-control points'' are identified in the core (such as glacial terminations, magnetic reversals, etc), and then ages for all the measurements are inferred from these control points, while accounting for compression using an involved heuristic process. 
Another  common approach is to align features of the $\delta^{18}$O time series to aspects of the
astronomical forcing, a process known as astronomical tuning.
For example, in \citet{Lisiecki2005}, ice ages are aligned with the predictions of the Imbrie and Imbrie (1980) model (linear relaxation  with different relaxation times for glaciation and deglaciation forced  by summer solstice insolation at 60$^{\circ}$ N).

The result of fitting an age model is a dataset $\left\{ \tau_m, Y_m\right\}_{m=1}^M$ in which $Y_m$ denotes the measurement of $\delta^{18}$O at time $\tau_m$.  It is widely known that dating estimates from age models, the $\{\tau_m\}$, are highly
uncertain, with accuracy believed to be of the order of 10 kyr
\citep{Huybers2007, Lisiecki2005}.
Investigating age 
models is beyond the scope of this paper, so we take as a starting point a stack  which
has been dated by other authors, and  we treat
$\tau_m$ as a given.  We reflect on the wisdom of this approach in Section \ref{S:Conclusion}.

In this article, we use 
the ODP677 record \citep{Shackleton1990}, shown in Figure~\ref{Fig:ModelComp}.
The foraminifera  here are of the benthic form, living in the deep ocean and therefore thought to be better representative of continental ice volume variations \citep[see, though][]{Elderfield12aa}. 
Analysis of other datasets can be found in \citet{Carson2014}. 
ODP677 has been dated both as part of an orbitally tuned scheme \citep{Lisiecki2005}, and a non-orbitally tuned scheme \citep{Huybers2007}, giving two different age estimates. 
We will refer to the data from the orbitally tuned scheme as ODP677-f, where the `f' denotes {\it forced}, and from the non-orbitally tuned scheme as ODP677-u, where the `u' denotes {\it unforced}. 
We focus on the last 780 kyr of this record (the last magnetic reversal occurred 780 kya, allowing us to date the starting point accurately), which contains 363 observations, and use it to highlight issues surrounding double counting of the astronomical forcing.

\subsection{The astronomical forcing}
\label{SS:AF}

The amount of insolation hitting the top of the atmosphere at any point on Earth 
is a function of the hour angle (time in the day), the latitude, and the true solar longitude (i.e., time in the year).
It also depends on  the obliquity, precession and eccentricity  of the Earth's orbit around the Sun, which vary over much longer time scales.
 {\it Obliquity} refers to the angle
between the equator and the orbital plane, and controls the seasonal contrast.   {\it Precession} of the point of
perihelion (the point of the orbit when the Earth is closest to the Sun) with respect to  the vernal point
marking the Spring equinox is quantified by the angle $\varpi$ made by the two points about the sun. It determines when in the seasonal cycle the Earth is closest to the Sun,  and  causes the positive/negative anomaly insolation patterns sequentially across the
different months of the year, thus controlling the length of the seasons.  {\it Eccentricity}, $e$, measures how much the Earth's orbit deviates from being circular and modulates the effect of precession.
Palaeoclimatologists often  transform eccentricity and precession, and refer instead to the {\it climatic precession}, $e\sin\varpi$, which is proxy for the effect of precession on the summer insolation in the Northern Hemisphere. By complementing climatic precession with  $e\cos\varpi$ (proxy for spring insolation, termed here {\it coprecession}), insolation may effectively be computed at any time in the year and for any latitude \citep{Berger1978a}. 

However, for  phenomenological models which are not typically spatially resolved, the practice is to use  
 some subset summary of the seasonal and spatial distribution of insolation.
For example,  Milankovitch theory (\citet{Milankovitch41_original}; translation in \citet{Milankovitch41}) asserts that the growth and shrinkage of ice sheets is controlled by summer insolation,
typically at a reference latitude of 60$^{\circ}$ N, a quantity 
 Milankovitch termed the caloric summer insolation.
  Other common summaries include the daily-mean insolation at summer solstice at 60$^{\circ}$ N \citep{Imbrie80},
or at different times in the year \citep{Saltzman1990}. 
These summaries may all be approximated as a linear combination of astronomical quantities as follows:

\begin{equation*}
F(t; \pmb{\gamma}) = \gamma_P \Pi_P (t) + \gamma_C \Pi_C (t) + \gamma_E E (t),
\end{equation*}
\noindent where $\Pi_P (t)$, $\Pi_C (t)$, and $E (t)$, are the normalised climatic precession, coprecession ($e\cos\varpi$), and obliquity respectively. The parameter $\pmb{\gamma}=(\gamma_P, \gamma_C, \gamma_E)^\top$ controls the linear combination. An algorithm to compute these quantities with sufficient accuracy for the late Pleistocene is provided in \citet{Berger1978a}. More accurate, time indexed data are provided by \citet{Laskar04aa} but the gain in accuracy is not critical in this context.

The geometry of ice sheets and snow line suggest  that a positive insolation anomaly may lead to a greater   ice volume change, than a negative one \citep{Ruddiman06aa}. To account for this,
some authors  truncate  the astronomical forcing  to down-weight
negative anomalies. Here, we introduce the truncation operator 
\[ f(x) = \left\{
  \begin{array}{l l}
    x + \sqrt{4a^2 + x^2} - 2a & \quad \text{if $x \leq 0$}\\
    x & \quad \text{otherwise,}
  \end{array} \right.
\]
which is used in model PP12 defined below \citep{paillard98,Parrenin2012}.

\subsection{Phenomenological models of climate dynamics}
\label{SS:Models}

We consider three models of the climate dynamics. They were each originally proposed as low-order ordinary differential equations, with state vector $\pmb{X}(t)=\left(X_{(1)}(t), ..., X_{(d)}(t) \right)^\top$, where $d$ is the dimension of the model, with the first component $X_{(1)}$ representing  global ice volume. The other
components represent quantities such as glaciation state, or CO$_2$ concentration. In order to account for model errors, we convert the models into stochastic differential equations by the addition of a Brownian motion $\pmb{W} (t)$.  These models were chosen as each  models the glacial--interglacial cycle
using a qualitatively different dynamical mechanism, as explained further below. For notational convenience we drop the explicit dependence of $\pmb{X}$ and $\pmb{W}$ on $t$. 

%For an overview of oscillators in palaeoclimate modelling see \cite{Crucifix2012}).

\subsubsection*{Model SM91: \citep{Saltzman1991}}
SM91 models glacial--interglacial cycles as a system of three SDEs,
\begin{eqnarray*}
dX_{(1)} & = & -\left( X_{(1)} + X_{(2)} + v X_{(3)} + F(\gamma_P,\gamma_C,\gamma_E) \right)dt + \sigma_1 dW_{(1)} \\
dX_{(2)} & = & \left( r X_{(2)} - p X_{(3)} - s X_{(2)}^2 - X_{(2)}^3 \right)dt + \sigma_2 dW_{(2)} \\
dX_{(3)} & = & -q \left( X_{(1)} + X_{(3)} \right) dt + \sigma_3 dW_{(3)}
\end{eqnarray*}
in which variables $X_{(2)}$ and $X_{(3)}$ represent 
CO2 concentration in the atmosphere  and deep-sea ocean temperature, respectively.  
The model exhibits limit-cycle dynamics,  oscillating with  a periodicity of around 100 kyr, although this is also  controlled by the astronomical forcing. 
The model is an expression of the hypothesis that
carbon-cycle effects are critical for the emergence of glacial cycles.
Hence the non-linear terms, which are responsible for the oscillation, are present in the second equation only. 
SM91 is non-dimensional with a reference value of 10 kyr for $t$.

\subsubsection*{Model T06: \citep{Tziperman2006} }
T06 is an example of a ``hybrid'' model coupling $X_{(1)}$, which is governed by a stochastic differential equation, 
to a binary indicator variable $X_{(2)}$.
\begin{eqnarray*}
dX_{(1)} & = & \left( \left(p_0 - K X_{(1)} \right) \left(1 - \alpha X_{(2)} \right) - \left(s + F(\gamma_P,\gamma_C,\gamma_E) \right) \right) dt + \sigma_1 dW_{(1)} \\
  X_{(2)} & : & \mbox{ switches from } 0 \mbox{ to } 1 \mbox{              when $X_{(1)}$ exceeds some threshold $T_u$} \\
  X_{(2)} & : & \mbox{ switches from } 1 \mbox{ to } 0 \mbox{              when $X_{(1)}$ falls below some threshold $T_l$}
\end{eqnarray*}
When $T_u$ and $T_l$ are suitably chosen, the resulting dynamics are that of a relaxation
oscillation: $X_{(1)}$ tends either to increase or decrease depending on the state of $X_{(2)}$, and
the trend reverses as $X_{(1)}$ crosses a threshold causing $X_{(2)}$ to switch state. As in SM91, the oscillation is further controlled by the astronomical forcing, allowing for synchronisation effects \citep{Tziperman2006}. 
The original motivation for T06 was to suggest a critical role for Arctic sea-ice cover \citep{Tziperman2004}; a positive sea-ice cover anomaly reduces the amount of snow fall on Northern Hemisphere ice caps, thereby acting \textit{negatively} on the growth of ice (hence the $- X_{(2)}$ term in the equation for $dX_{(1)}$), and vice-versa. 
 Ice volume is expressed in units of $10^{15} \mbox{m}^3$, $K$ in units of $\mbox{kyr}^{-1}$, and $p_0$ and $s$ in units of $10^6 \mbox{m}^3 \mbox{s}^{-1}$.

\subsubsection*{Model PP12: \citep{Parrenin2012}}
PP12 is also a hybrid model,  with 
 $X_{(2)}$ now
  representing a hidden climatic state, that may either be ``glaciation" (0) or ``deglaciation" (1).
\begin{eqnarray*}
  dX_{(1)} &= &-(\gamma_P \Pi_P^{\dag} + \gamma_C \Pi_C^{\dag} + \gamma_E E - a_g + (a_g+a_d+X_{(1)}/\tau)X_{(2)})dt \\
 & & + \sigma_1 dW_{(1)}, \\ 
  X_{(2)} & : & \mbox{ switches from } 0 \mbox{ to } 1 \mbox{ when $F(\kappa_P,\kappa_C,\kappa_E)$ is less than some }\\
  & & \mbox{ threshold $v_\text{l}$} \\
X_{(2)} & : & \mbox{ switches from } 1 \mbox{ to } 0 \mbox{ when  $F(\kappa_P,\kappa_C,\kappa_E) + X_{(1)}$ is greater }
\\ & & \mbox{ than some threshold $v_\text{u}$}
\end{eqnarray*} 
where $\Pi^{\dag}$ and $\varPi^{\dag}$ are transformed precession and coprecession components defined to be
\begin{eqnarray*}
\Pi_P^{\dag} & = & (f(\Pi_P) - 0.148) \diagup 0.808 \\
\Pi_C^{\dag} & = & (f(\Pi_C) - 0.148) \diagup 0.808.
\end{eqnarray*}
 This discrete distinction of 
  climatic states is given an empirical justification in the data analysis by \citet{Imbrie11aa}. 
  During the glaciation phase, ice volume trends upwards and is linearly controlled by the truncated insolation. 
  The deglaciation phase is simply a relaxation towards low ice volume. The model assumes that 
  the switch from deglaciation to glaciation is controlled by insolation alone, 
  whereas the glaciation-deglaciation switch is determined by a condition on glaciation and insolation.  This contrasts with T06, where the 
  state changes are determined only by the system state. Consequently, a constant astronomical forcing cannot induce spontaneous oscillations in PP12. Ice volume is expressed as sea-level  equivalent in meters, $\gamma_P$, $\gamma_C$, $\gamma_E$, $a_g$, and $a_d$ in units of $\mbox{m(kyr)$^{-1}$}$, $\tau$ in kyr, and $\kappa_P$, $\kappa_C$, $\kappa_E$, $v_l$, and $v_u$ in meters.  \\

\begin{figure}
\centering
\includegraphics[width=\textwidth]{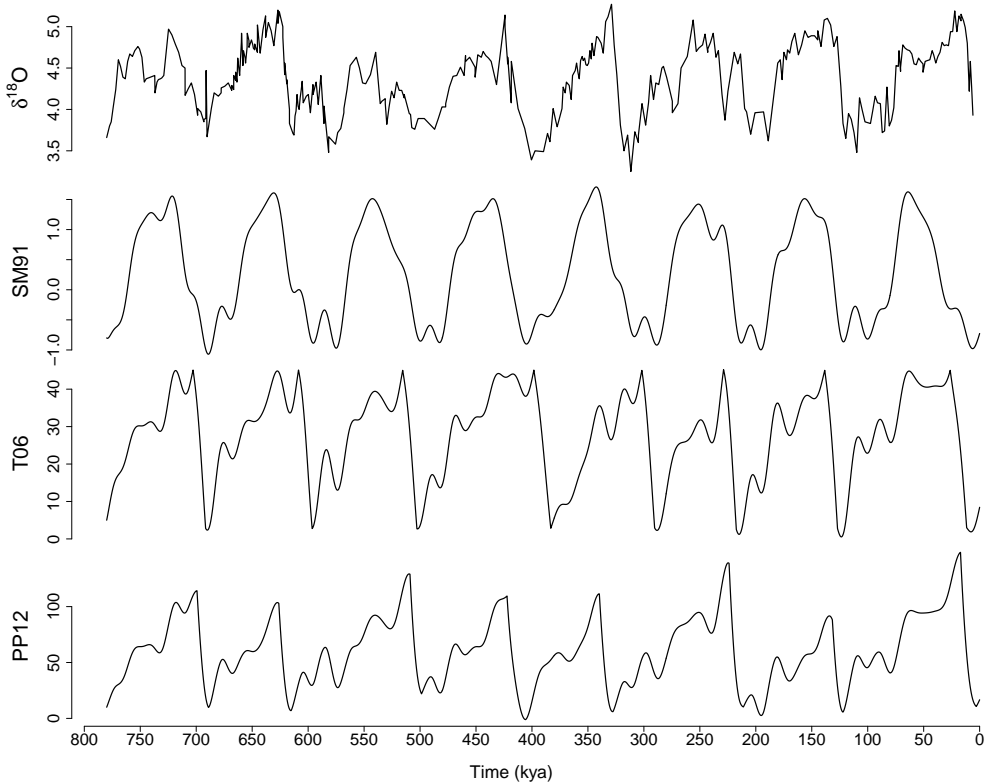}
\caption{Top row: Observed $\delta^{18}\mbox{O}$ from ODP677 \citep{Shackleton1990} corresponding to the past 780 kyr with the H07 chronology \citep{Huybers2007}. Rows 2-4: Comparison of the ice volume generated from deterministic versions of SM91, T06, and PP12.}
\label{Fig:ModelComp}
\end{figure}

\noindent A comparison of the ice volume generated from the deterministic version of each model is shown in Figure~\ref{Fig:ModelComp}, using the parameters suggested in the original publications. Each model captures the broad structure of the glacial-interglacial cycle. 
These plots are not precise reproductions of the  figures  in the original publications, due to differences in the initial conditions and astronomical solutions.
Note that each  model was tuned using a different dataset, and so, for example, SM91 has seven cycles in 780 kyr rather than eight. 

\subsection{Statistical observation model}
\label{SS:Obs}

The final modelling ingredient is a statistical model relating the  state
variables in the dynamical climate models,  $\pmb{X} (t)$, to the  dataset.  We assume that the data are of the form $\left\{ \tau_m, Y_m\right\}_{m=1}^M$, where $\tau_m$ is the estimated age and $Y_m$ the measured 
proxy of the $m^{th}$ data point/slice. 
We use the model
\[ Y_m \sim \mathcal{N} ( D + \pmb{H}^\top \pmb{X}_m, \sigma_y^2), \]
\noindent where we  define $\pmb{X}_m = \pmb{X} (\tau_m)$. Here, we use $\pmb{H}=(C, 0,\ldots, 0)^\top$, so that $Y_m$ is a scaled and shifted version of the value $X_{(1)} (\tau_m)$, the ice volume in the underlying dynamical model. However, vector observations can be used at no additional cost or complication to the methodology, allowing observations of other proxies if desired.

\section{Methodology}
\label{S:Meth}

Our primary aim is model comparison, in particular to answer the question: given a collection of competing models $\{\mathcal{M}_l\}_{l=1}^L$, which is best supported by the data? %The Bayesian approach to model selection uses Bayes factors (BF) \cite{Jeffreys1939, Kass1995}. 
The Bayes factor (BF) for comparing two models, $\mathcal{M}_1$ and $\mathcal{M}_2$ say, is the ratio of their evidences
\begin{equation*}
B_{12} = \frac{\pi(Y_{1:M} \mid \mathcal{M}_1)}{\pi(Y_{1:M} \mid \mathcal{M}_2)},
\end{equation*}
\noindent where $Y_{1:M} = (Y_1, \ldots, Y_M)$ and where $\pi(Y_{1:M} \mid \mathcal{M}_l)$ is the evidence for model $\mathcal{M}_l$ \citep{Jeffreys1939, Kass1995}. The Bayes factor summarises the strength of evidence in the data in support of one model over another, and is the ratio of the posterior to the prior odds in favour of $\mathcal{M}_1$ over $\mathcal{M}_2$. If the prior probabilities for each model are equal, then the Bayes factor is the ratio of the posterior model probabilities.

Secondary aims of our analysis include parameter estimation and filtering, which in this context are often called calibration and climate reconstruction (or hindcasting). 
Calibration is the process of finding the posterior distribution of the model parameters $\pi(\pmb{\theta}_l \mid Y_{1:M}, \mathcal{M}_l)$,  where $\pmb{\theta}_l$ is the parameter for model $\mathcal{M}_l$, and filtering is finding the distribution of the state variables $\pi(\pmb{X}_{1:M} \mid Y_{1:M}, \pmb{\theta}_l, \mathcal{M}_l)$. These three problems are of different levels of difficulty. Filtering is the most straightforward, but is not simple as for non-linear or non-Gaussian models, direct calculation of the filtering distributions is  not possible, and so we must instead rely upon approximations.
Calibration requires that we integrate out the dependence on the state variables $\pmb{X}_{1:M}$, 
\begin{equation*}
\pi (\pmb{\theta}_l \mid Y_{1:M}, \mathcal{M}_l) = \int \pi(\pmb{\theta}_l, \pmb{X}_{1:M} \mid Y_{1:M}, \mathcal{M}_l) {\rm d}\pmb{X}_{1:M},
\end{equation*}
and hence, is considerably more difficult than filtering. 
Finally, model selection requires integrating out the dependence on $\pmb{\theta}_l$, 
\begin{multline*}
\pi (Y_{1:M} \mid \mathcal{M}_l) = \int \pi(\pmb{\theta}_l \mid \mathcal{M}_l) \int \pi \left(\pmb{X}_{1:M} \mid \pmb{\theta}_l, \mathcal{M}_l \right) \times \\ \pi(Y_{1:M} \mid \pmb{\theta}_l, \pmb{X}_{1:M}, \mathcal{M}_l) {\rm d}\pmb{X}_{1:M} {\rm d}\pmb{\theta}_l,
\end{multline*}
and is thus even more difficult than calibration. 

The development of Monte Carlo methodology for solving these three problems for state space models reflects this hierarchy of difficulty. Particle filter methodology, first proposed in the 1990s \citep{Gordon1993}, is able to solve the general filtering problem adequately as long as the dimension of $\pmb{X}_{1:M}$ is not too large. Whereas  the calibration problem   has  only begun to be satisfactorily answered more recently, with the development of pseudo-marginal methods such as particle-MCMC \citep{Andrieu2010}. Calculating the model evidence is, however, still very much an open problem.

Here, we demonstrate how the recently introduced SMC$^2$ algorithm \citep{Chopin2012} can be used to estimate model evidences.
The approach relies upon the following identities decomposing the evidence:
\begin{equation}\label{eqn:Ident1}
\pi(Y_{1:M}) =\pi(Y_1) \prod_{m=2}^M \pi(Y_m \mid Y_{1:m-1}),
\end{equation}
and 
\begin{equation}\label{eqn:Ident2}
\pi(Y_m \mid Y_{1:m-1}) = \int \pi(Y_m \mid Y_{1:m-1}, \pmb{\theta}) \pi(\pmb{\theta} \mid Y_{1:m-1}) {\rm d} \pmb{\theta},
\end{equation}
where we have dropped the dependence on $\mathcal{M}_l$ from the notation.
SMC$^2$ can be used to sample from $\pi(\pmb{\theta} \mid Y_{1:m-1})$ and find unbiased estimates of $ \pi(Y_m \mid Y_{1:m-1}, \pmb{\theta})$, and  remarkably, it can be shown that plugging these estimates into Equations (\ref{eqn:Ident1}) and (\ref{eqn:Ident2}) gives an  unbiased estimate of the model evidence \citep{Andrieu2009}. Note, however, that the ratio of two unbiased model evidence estimates gives a consistent but biased estimator of the BF.

\subsection{Estimating Model Evidence Using SMC$^2$}
\label{SS:SMC2}

Sequential Monte Carlo algorithms (SMC) \citep{DelMoral2006} are population based sampling methods that aim to  sample from some target distribution, $\pi_M$, by sampling from a  series of intermediary distributions, $\{ \pi_m \}_{m=1}^M$, that  are chosen to gradually  `close-in' on the target distribution. SMC uses a weighted collection of particles to approximate each distribution, and sequentially updates the weights and the particles in such a way that the normalising constant of each distribution can be estimated.
A common choice for the sequence of distributions is to add a single data point at a time, so that the intermediary distributions are $\pi(\pmb{\theta} \mid Y_{1:m})$ or $\pi(\pmb{X}_{1:m} \mid Y_{1:m})$, for example.

One of the earliest SMC algorithms is the particle filter (PF) \citep{Gordon1993}, which  samples from the sequence of filtering distributions $\pi_m(\pmb{X}_{1:m}) = \pi(\pmb{X}_{1:m} \mid Y_{1:m}, \pmb{\theta})$, and is described in  Algorithm~\ref{A:PF}. The basic idea is that at  initialisation, a sample of $N_x$ particles are sampled from some initial proposal density $r_1(\pmb{X}_1 \mid Y_1, \pmb{\theta})$, and given importance weight $\pi (\pmb{X}_1, Y_1 \mid \pmb{\theta}) \diagup r_1(\pmb{X}_1 \mid Y_1, \pmb{\theta})$.
These particles are then repeatedly resampled, propagated via some arbitrary proposal distribution $r_m(\pmb{X}_m \mid \pmb{X}_{1:m}, Y_{1:m}, \pmb{\theta})$, and reweighted accordingly, so that for each successive iteration the particles are a weighted sample of the posterior $\pi(\pmb{X}_{1:m} \mid Y_{1:m}, \pmb{\theta})$.
Details of the resampling  and the proposal distributions, $r_m$, are discussed in Section \ref{SS:Proposals}, and further details can be found in \citet{Doucet2009}.

\begin{algorithm}[p]

\caption{Particle filter targeting $\pi \left(\pmb{X}_{1:M} \mid Y_{1:M}, \pmb{\theta} \right)$.}
\label{A:PF}

\begin{algorithmic}

\FOR{$k=1,...,N_X$}
\STATE{Sample $\pmb{X}_1^{\left(k \right)} \sim r_1\left(\pmb{X}_1 \mid Y_1, \pmb{\theta} \right)$.}
\STATE{Set the importance weight
\[ \omega_1^{(k)} = \frac{\pi \left(\pmb{X}_1^{\left(k \right)} \mid \pmb{\theta} \right) \pi \left(Y_1 \mid \pmb{X}_1^{\left(k \right)}, \pmb{\theta} \right)}{r_1\left(\pmb{X}_1^{\left(k \right)} \mid Y_1,\pmb{\theta} \right)}. \]}
\ENDFOR
\STATE{Normalise the weights. For $k=1,...,N_X$
\[ \Omega_1^{(k)} = \frac{\omega_1^{(k)}}{\sum_{i=1}^{N_X} \omega_1^{(i)}}. \]}
\FOR{$m=2,...,M$}
\FOR{$k=1,...,N_X$}
\STATE{Sample ancestor particle index $a_{m-1}^{(k)}$ according to weights $\Omega_{m-1}^{(1:N_X)}$.}
\STATE{Sample $\pmb{X}_m^{\left(k \right)} \sim r_m\left(\pmb{X}_m \mid \pmb{X}_{m-1}^{\left(a_{m-1}^{(k)} \right)} , Y_m, \pmb{\theta} \right)$.}
\STATE{Extend particle trajectory $\pmb{X}_{1:m}^{(k)} = \left\lbrace \pmb{X}_{1:m-1}^{(a_{m-1}^{(k)})}, \pmb{X}_m^{\left(k \right)} \right\rbrace$.}
\STATE{Set the importance weight
\begin{equation} \omega_m^{(k)} = \frac{\pi \left(\pmb{X}_m^{\left(k \right)} \mid \pmb{X}_{m-1}^{\left(a_{m-1}^{(k)} \right)}, \pmb{\theta} \right) \pi \left( Y_m \mid \pmb{X}_m^{\left(k \right)}, \pmb{\theta} \right)}{r_m\left(\pmb{X}_m^{\left(k \right)} \mid \pmb{X}_{m-1}^{\left(a_{m-1}^{(k)} \right)} , Y_m, \pmb{\theta} \right)}. \label{eqn:ISweight}\end{equation}}
\ENDFOR
\STATE{Normalise the weights. For $k=1,...,N_X$
\[ \Omega_m^{(k)} = \frac{\omega_m^{(k)}}{\sum_{i=1}^{N_X} \omega_m^{(i)}}. \]}
\ENDFOR
\end{algorithmic}

\end{algorithm}

An important aspect of the PF is that an unbiased estimate of the normalising constant $\pi(Y_{1:m} \mid \pmb{\theta})$ can be estimated from the unnormalised weights in each iteration of the algorithm, using 
\begin{equation*}
\hat{\pi} (Y_m \mid Y_{1:m-1}, \pmb{\theta}) = \frac{1}{N_x} \sum_{k=1}^{N_x} \omega_m^{(k)} (\pmb{X}_{1:m}^{(k)})
\end{equation*}
as an approximation to Equation (\ref{eqn:Ident2}), and then
plugging these estimates into Equation (\ref{eqn:Ident1})  \citep{DelMoral2004}
\begin{equation}\label{eqn:Ident3}
\hat{\pi} (Y_{1:M} \mid \pmb{\theta}) = \hat{\pi}(Y_1) \prod_{m=2}^M \hat{\pi} (Y_m \mid Y_{1:m-1}, \pmb{\theta}).
\end{equation}
In \citet{Andrieu2009}, it was shown that using these unbiased estimates of the likelihood in other Monte Carlo algorithms can lead to valid Monte Carlo algorithms (termed pseudo-marginal algorithms) for performing parameter estimation. For example, PMCMC \citep{Andrieu2010} uses the PF within an MCMC algorithm, and SMC$^2$ \citep{Chopin2012} uses a PF embedded within an SMC algorithm, both with the aim of finding $\pi(\pmb{\theta} \mid Y_{1:M})$. We concentrate on the latter as it allows for estimation of BFs.

The SMC$^2$ algorithm \citep{Chopin2012} embeds the particle filter within an SMC algorithm targeting the sequence of posteriors
\begin{equation*}
\pi_0 = \pi(\pmb{\theta}), \hspace{1cm} \pi_m = \pi(\pmb{\theta}, \pmb{X}_{1:m} \mid Y_{1:m}),
\end{equation*}
for $m=1, \ldots, M$. 
This is achieved by initially sampling $N_\theta$ parameter particles, $\{\pmb{\theta}^{(n)}\}_{n=1}^{N_\theta}$ from the prior. To each $\pmb{\theta}^{(n)}$, we attach a PF of $N_x$ particles, i.e., at iteration $m$ the PF $\{\pmb{X}_{1:m}^{(k,n)}, \Omega_m^{(k,n)}\}_{k=1}^{N_x}$ is associated with $\pmb{\theta}^{(n)}$,  where  $\Omega_m^{(k,n)}$ are the normalised weights in Algorithm \ref{A:PF}. From this PF we can obtain
 an unbiased estimate of the marginal likelihood $\pi(Y_{1:m} \mid\pmb{\theta}^{(n)})$ via Equation (\ref{eqn:Ident3}). To assimilate the next observation $Y_{m+1}$, we first extend the PF for the X-states to $\{\pmb{X}_{1:m+1}^{(k,n)}, \Omega_{m+1}^{(k,n)}\}_{k=1}^{N_x}$, and then estimate $\pi(Y_{1:m+1} \mid \pmb{\theta}^{(n)})$ and so on. Particle
degeneracy occurs when the weighted particle approximation is dominated by just  a few particles (i.e., a few have comparatively large weights), and is monitored by calculating the effective sample size (ESS) 
\[\mbox{ESS } = \left( \sum_{i=1}^{N_\theta}  \left( W_m^{(i)} \right)^2 \right)^{-1}, \] 
\noindent where $\left\lbrace W_m^{(i)} \right\rbrace_{i=1}^{N_\theta}$ are the normalised weights in population $m$.
When the ESS falls below some threshold (usually $N_\theta / 2$) the particles are resampled to discard low-weight particles.
However, resampling can lead to too few unique particles in the parameter space.
Particle diversity is improved by running a PMCMC algorithm that leaves $\pi (\pmb{\theta}, \pmb{X}_{1:m} \mid Y_{1:m})$ invariant, specifically the particle marginal Metropolis-Hastings (PMMH) algorithm \citep{Andrieu2010}.
The full details of the SMC$^2$ algorithm are presented in Algorithm~\ref{A:SMC2M}, with  theoretical justification in \citet{Chopin2012}.

\begin{algorithm}[p]

\caption{SMC$^2$ algorithm targeting $\pi \left( \theta, X_{1:M} \mid Y_{1:M} \right)$.}
\label{A:SMC2M}

\begin{algorithmic}

\FOR{$n=1,...,N_{\theta}$}
\STATE{Sample $\pmb{\theta}^{(n)}$ from the prior distribution, $\pi \left(\pmb{\theta}\right)$.}
\STATE{Set the importance weight $W^{(n)}_0 = 1 \diagup N_\theta$.}
\ENDFOR

\FOR{$m=1,...,M$}
\IF{ESS$<\frac{N_{\theta}}{2}$}
\FOR{$n=1,...,N_{\theta}$}
\STATE{Sample $\pmb{\theta}^{*(n)}$ and $\pmb{X}_{1:m-1}^{*(1:N_X,n)}$ from $\pmb{\theta}^{(1:N_{\theta})}$ and $\pmb{X}_{1:m-1}^{(1:N_X,1:N_{\theta})}$, according to weights $W_{m-1}^{(1:N_{\theta})}$.}
\STATE{Sample $\pmb{\theta}^{**(n)}$ and $\pmb{X}_{1:m-1}^{**(1:N_X,n)}$ from a PMMH algorithm targeting  $\pi \left(\pmb{\theta}, \pmb{X}_{1:m-1} \mid Y_{1:m-1} \right)$ initialised with $\pmb{\theta}^{*(n)}$ and $\pmb{X}_{1:m-1}^{*(1:N_X,n)}$.}
\ENDFOR 
\STATE{Set $\pmb{\theta}^{(1:N_\theta)} = \pmb{\theta}^{**(1:N_\theta)}$ and $\pmb{X}_{1:m-1}^{(1:N_X,1:N_\theta)} = \pmb{X}_{1:m-1}^{**(1:N_X,1:N_\theta)}$.}
\STATE{Set the importance weights $W_{m-1}^{(n)}=1 \diagup N_{\theta} \mbox{ for } n=1,...,n_\theta$.}
\ENDIF 
\FOR{$n=1,...,N_{\theta}$}
\STATE{Sample $\pmb{X}_{1:m}^{(1:N_X,n)}$ by performing iteration $m$ of the particle filter, and record estimates of $\hat{\pi} \left( Y_m \mid Y_{1:m-1}, \pmb{\theta}^{(n)} \right)$ and $\hat{\pi} \left( Y_{1:m} \mid \pmb{\theta}^{(n)} \right)$.}
\STATE{Set the importance weights $w^{(n)}_m = w^{(n)}_{m-1} \hat{\pi} \left( Y_m \mid Y_{1:m-1}, \pmb{\theta}^{(n)} \right)$. }
\ENDFOR
\STATE{Normalise the weights  \[ W^{(n)}_m = \frac{w^{(n)}_m}{\sum_{i=1}^{N_\theta} w^{(i)}_m}\; \mbox{ for }\; n=1,...,N_\theta. \]}
\STATE{Evaluate \[ \hat{\pi} \left(Y_m \mid Y_{1:m-1} \right) = \sum_{i=1}^{N_\theta} W_m^{(i)} \hat{\pi} \left(Y_m \mid Y_{1:m-1}, \pmb{\theta}^{(i)} \right).\]}
\ENDFOR

\end{algorithmic}

\end{algorithm}

The model evidence $\pi(Y_{1:M})$ can be decomposed according to Equation (\ref{eqn:Ident1}), and in each iteration of the SMC$^2$ algorithm, the term 
\[ \hat{\pi}(Y_m \mid Y_{m-1}) = \sum_{n=1}^{N_\theta} W^{(n)} \hat{\pi} (Y_m \mid Y_{1:m-1}, \pmb{\theta}^{(n)}) \]
provides an unbiased estimate of $\pi(Y_m \mid Y_{m-1})$.
An unbiased estimate of the model evidence $\pi(Y_{1:M})$ is then obtained from (\ref{eqn:Ident1}) with 
${\hat \pi}(Y_m \mid Y_{m-1})$ substituted for ${\pi}(Y_m \mid Y_{m-1})$. 

\subsection{Guided proposals}
\label{SS:Proposals}

A further difficulty arises as the transition densities 
$\pi(\pmb{X}_m \mid \pmb{X}_{m-1}, \pmb{\theta})$ are not available in closed form for the models of interest, 
suggesting that we need to choose the particle proposal distributions, $\{r_m\}$, so that the transition density cancels from the importance weights. 
This can be achieved by setting $r_m= \pi(\pmb{X}_m \mid \pmb{X}_{m-1}, \pmb{\theta})$ in Equation (\ref{eqn:ISweight}), so that proposals are just simulations from the model.
However, this choice will typically lead to particle degeneracy if too many of the proposals end up being far from the observations, due to the light Gaussian tails in the observation model.
Resampling the state particles ensures that important particles are propagated forward, which can improve the approximation in later iterations. Multinomial resampling is the most commonly used resampling scheme, but  alternatives such as stratified resampling give improvements in  sample variance \citep{Liu1998,Douc2005}. 

For the SDE models considered here, ressampling is not sufficient to overcome the  degeneracy problem. Our solution
is to avoid using  the model as the proposal distribution, and to instead build novel Brownian bridge type proposals, based on the proposals developed in \citet{Golightly2008},
 that guide the particles toward the next data point (thus decreasing degeneracy).
The key is to exploit the  Euler-Maruyama approximation we use to simulate from the underlying SDE, in order to condition the proposal distribution on the next observation, thus increasing the number of proposals with large weights. 
Each of our models are SDEs of the form
\begin{equation*}
d\pmb{X} \left(t\right) = \pmb{\mu} \left(\pmb{X}\left(t\right), \pmb{\theta} \right) dt + \mathit{\Sigma}_X^{\frac{1}{2}} \left(\pmb{X}\left(t\right), \pmb{\theta} \right) d\pmb{W}\left(t\right).
\end{equation*}
The Euler-Maruyama approximation simulates from the SDE over time interval $\Delta t$
 by partitioning the  interval into $J$ sub-intervals of length $\delta t = \frac{\Delta t}{J}$, and using  the discrete time equation 
\begin{equation*}
\pmb{X} \left(t' + \delta t \right) = \pmb{\mu} \left(\pmb{X}\left(t'\right), \pmb{\theta} \right) \delta t + \mathit{\Sigma}_X^{\frac{1}{2}} \left(\pmb{X}\left(t'\right), \pmb{\theta} \right) {\delta t}^{\frac{1}{2}} \pmb{\epsilon}_t,
\end{equation*}
\noindent where $\pmb{\epsilon}_t$ is a vector of independent standard Gaussian random variables.  Simulating from the discrete time equation between two observation times, $\tau_{m}$ and $\tau_{m+1}$, introduces $\left(J-1\right) \times d$ latent variables, $\pmb{X}_{m-1,1},...,\pmb{X}_{m-1,J-1}$, where we let $\pmb{X}_{m,j} = \pmb{X}\left( t_m + j \cdot \delta t \right)$. We can extend the particle filter to also sample from these latent variables, by using a proposal distribution of the form $\tilde{r}_{m+1} \left( \pmb{X}_{m,1},...,\pmb{X}_{m,J} \mid Y_{m+1}, \pmb{X}_m, \pmb{\theta} \right)$. The importance weight calculation in the particle filter is then
\begin{equation*}
\omega_{m+1}^{(k)} = \frac{\prod_{j=1}^{J} \pi \left(\pmb{X}_{m,j} \mid \pmb{X}_{m,j-1}, \pmb{\theta} \right) \pi\left(Y_{m+1} \mid \pmb{X}_{m+1}, \pmb{\theta} \right)}{\tilde{r}_m \left( \pmb{X}_{m,1},...,\pmb{X}_{m,J} \mid Y_{m+1}, \pmb{\theta} \right)},
\end{equation*}
\noindent where the $\pi \left(\pmb{X}_{m,j} \mid \pmb{X}_{m,j-1}, \pmb{\theta} \right)$ are now assumed to be Gaussian densities. 

We can  guide the particles into regions of high likelihood by conditioning the value of $\pmb{X}_{m,j+1}$ on  future observation, $Y_{m+1}$. This can be done by approximating the distribution of $Y_{m+1}$ conditional on $\pmb{X}_{m,j}$ using a single Euler-Maruyama step of size $\widetilde{\Delta t} = \Delta t - j \delta t$. To do this conditioning, note that under an Euler-Maruyama step of interval size $\widetilde{\Delta t}$,
\begin{equation*}
\pmb{X}_{m+1} \mid \pmb{X}_{m,j}, \pmb{\theta} \sim \mathcal{N}_d \left( \pmb{X}_{m,j} + \pmb{\mu}_{m,j} \widetilde{\Delta t}, \mathit{\Sigma}_{m,j} \widetilde{\Delta t} \right),
\end{equation*}
where $\pmb{\mu}_{m,j} = \pmb{\mu} \left( \pmb{X}_{m,j}, \pmb{\theta} \right)$ and $\mathit{\Sigma}_{m,j} = \mathit{\Sigma}_X \left( \pmb{X}_{m,j}, \pmb{\theta} \right)$.
We can then see that the  joint distribution of $\pmb{X}_{m,j+1}$ and $Y_{m+1}$, given $\pmb{X}_{m,j}$, is 
\begin{multline*}
\left(\!\! \begin{array}{c} \pmb{X}_{m,j+1} \\ Y_{m+1} \end{array}\! \!\right) \mid \pmb{X}_{m,j}, \pmb{\theta} \sim \mathcal{N}_{d+1} \left(\mkern-6mu \left(\mkern-6mu \begin{array}{c} \pmb{X}_{m,j} + \pmb{\mu}_{m,j} \delta t \\ \pmb{H} \left( \pmb{X}_{m,j} + \pmb{\mu}_{m,j} \widetilde{\Delta t} \right) + D \end{array} \mkern-6mu\right),\mkern-6mu \right. \\ \left. \left( \mkern-6mu\begin{array}{c c} \mathit{\Sigma}_{m,j} \delta t  & \mathit{\Sigma}_{m,j} \pmb{H}^\top \delta t  \\ \pmb{H} \mathit{\Sigma}_{m,j} \delta t & \pmb{H} \mathit{\Sigma}_{m,j} \pmb{H}^\top \widetilde{\Delta t} + \sigma_y^2 \end{array} \mkern-6mu\right)  \mkern-6mu \right).
\end{multline*}

\noindent Conditioning this distribution on  $Y_m$ \citep{Eaton1983}, then suggests  proposals of the form 
\begin{equation*}
\pmb{X}_{m,j+1} \mid \pmb{X}_{m,j}, Y_{m+1}, \pmb{\theta} \sim  \mathcal{N}_d \left(\pmb{\mathcal{M}}_{m,j}, \pmb{\mathcal{S}}_{m,j} \right),
\end{equation*}
\noindent where
\begin{equation*}
\pmb{\mathcal{M}}_{m,j} =  \pmb{X}_{m,j}+\pmb{\mu}_{m,j} \delta t +  \pmb{B}^\top \pmb{A}^{-1} \left(Y_{m+1} - \pmb{H} \left( \pmb{X}_{m,j} + \pmb{\mu}_{m,j} \widetilde{\Delta t} \right) - D  \right),
\end{equation*}
\noindent and
\begin{equation*}
\pmb{\mathcal{S}}_{m,j} =  \mathit{\Sigma}_{m,j} \delta t - \pmb{B}^\top \pmb{A}^{-1} \pmb{B},
\end{equation*}
\noindent with
\begin{equation*}
\pmb{A} = \left(\pmb{H} \mathit{\Sigma}_{m,j} \pmb{H}^\top \widetilde{\Delta t} + \sigma_y^2\right)
\mbox{ and }
\pmb{B} = \pmb{H} \mathit{\Sigma}_{m,j} \delta t.
\end{equation*}
\noindent In our experiments, we have found that using these guided proposals dramatically reduces particle degeneracy. This improves the likelihood estimates, thus increasing the efficiency of the algorithm. Consequently, smaller value of $N_x$ (fewer $X$-particles) can be used, and the PMMH rejuvenation step has better mixing properties, allowing for shorter chains.

\subsection{Further details}
\label{SS:Det}

\noindent The tuning parameters are the number of particles, $N_\theta$ and $N_x$, and the proposal distributions for the PMMH rejuvenation steps. Typically, $N_\theta$ will be decided by the available computational resource. A low value of $N_x$ can be used for early iterations, but must be increased when using longer time series of data in later iterations.
An insufficient number of state particles  has a negative impact on the PMCMC acceptance rate, leading to fewer acceptances.
Automatic calibration of $N_x$ is discussed in \citet{Chopin2012}, where it is suggested that $N_x$ is doubled whenever the acceptance rate of the PMCMC step becomes too small.
 We use $N_x = N_\theta = 1000$ throughout.
The fact that we have a collection of particles in each iteration allows automated calibration of the PMMH proposals. For example, using the sample mean and variance to design a random-walk proposal, or using a Gaussian independence sampler.
We use independent Gaussian proposals using the sample mean and covariance, with a chain length of 10 to maintain a high particle diversity.

\section{Results}
\label{S:Results}

\subsection{Simulation study}
\label{SS:SS}
\begin{figure}
\centering
\includegraphics[width=\textwidth]{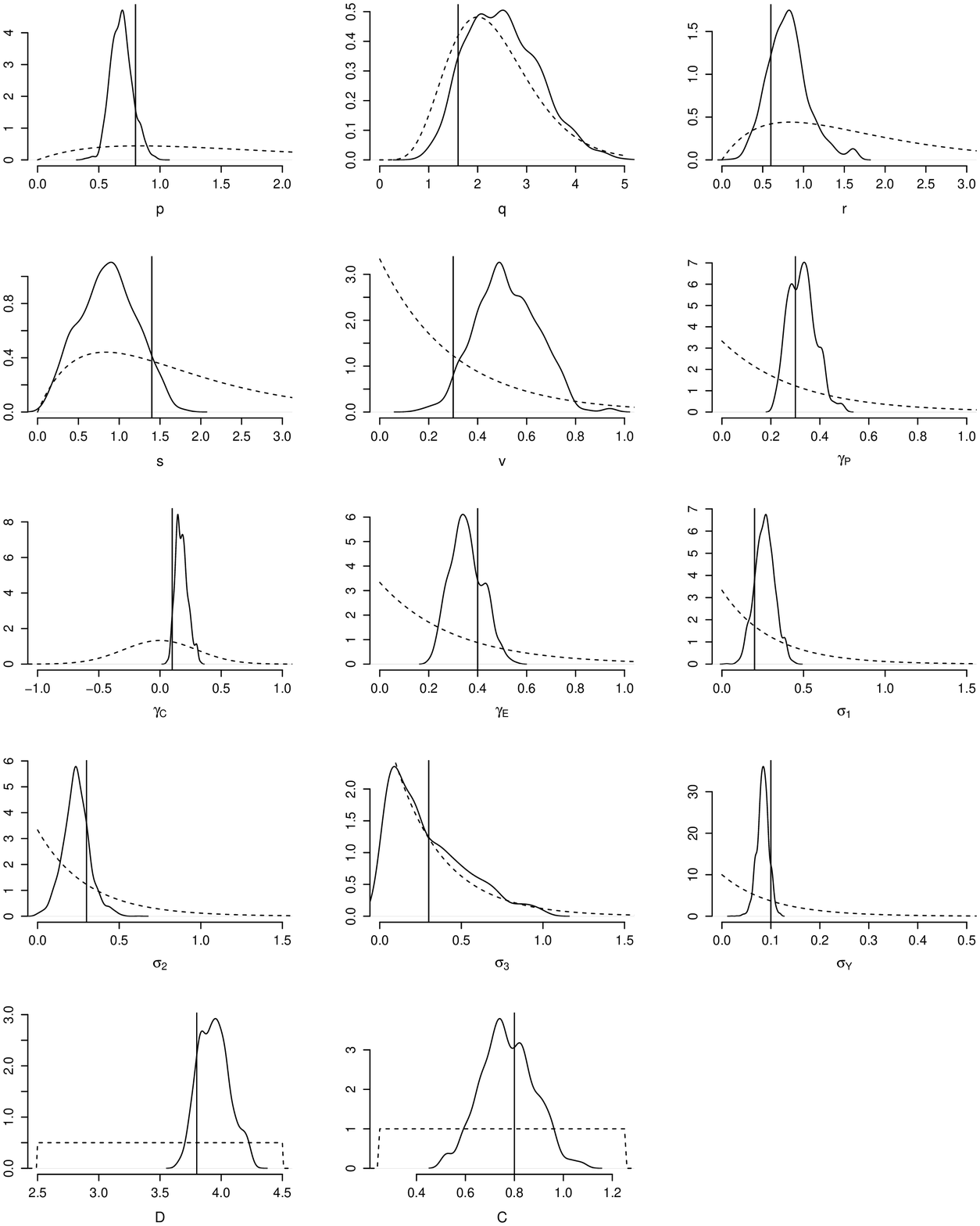}
\caption{
Marginal posterior distributions for the parameters of the forced SM91 model when fit to the  SM91-f dataset.
Vertical lines show the parameter values used to generate the data, and dashed lines represent the prior distribution. 
}
\label{Fig:PosteriorSS}
\end{figure}

In order to gain confidence in the ability of our SMC$^2$ algorithm for both model selection and calibration, we begin with a simulation study.
We simulate a single random trajectory from a given model and parameter setting and draw observations from the observation process. We then show that the posterior distributions recover the true value of the parameters (Figure \ref{Fig:PosteriorSS}), and that the Bayes factors correctly identify the true generative model (Table \ref{Tab:EvidenceSS}).

\begin{table}
\caption{\label{Tab:Priors}Prior distributions used for each model in both the simulation study and the analysis of ODP677.}
\centering
\begin{tabular}{*{3}{l}}
\hline
SM91 & T06 &PP12 \\
\hline
$\gamma_P \sim \operatorname{Exp}(1 \diagup 0.3)$ & $\gamma_P\sim \operatorname{Exp}(1 \diagup 0.6)$ & $\gamma_P\sim \operatorname{Exp}(1 \diagup 1.5)$ \\
$\gamma_C\sim \mathcal{N}(0, 0.3)$ & $\gamma_C\sim \mathcal{N}(0, 0.6)$ & $\gamma_C\sim \mathcal{N}(0, 1.5)$ \\
$\gamma_E\sim \operatorname{Exp}(1 \diagup 0.3)$ & $\gamma_E\sim \operatorname{Exp}(1 \diagup 0.6)$ & $\gamma_E\sim \operatorname{Exp}(1 \diagup 1.5)$ \\
&  & \\
$p\sim \Gamma(2,1.2)$ & $p_0\sim \operatorname{Exp}(1 \diagup 0.3)$ & $a\sim \Gamma(8,0.1)$\\
$q\sim \Gamma(7,3)$ & $K\sim \operatorname{Exp}(1 \diagup 0.1)$ & $a_d\sim \operatorname{Exp}(1)$\\
$r\sim \Gamma(2,1.2)$ & $s\sim \operatorname{Exp}(1 \diagup 0.3)$ & $a_g\sim \operatorname{Exp}(1)$\\
$s\sim \Gamma(2,1.2)$ & $\alpha\sim \operatorname{Beta}(40,30)$ & $\kappa_P\sim \operatorname{Exp}(1 \diagup 20)$ \\ 
$v\sim \operatorname{Exp}(1/0.3)$ & $x_l\sim \operatorname{Exp}(1 \diagup 3)$ & $\kappa_C\sim \mathcal{N}(0, 20)$ \\
$\sigma_1\sim \operatorname{Exp}(1 \diagup 0.3)$ & $x_u\sim \Gamma(90,0.5)$ & $\kappa_E\sim \operatorname{Exp}(1 \diagup 20)$ \\
$\sigma_2\sim \operatorname{Exp}(1 \diagup 0.3)$ & $\sigma_1\sim \operatorname{Exp}(1 \diagup 2)$ & $\tau\sim \operatorname{Exp}(1 \diagup 10)$ \\
$\sigma_3\sim \operatorname{Exp}(1 \diagup 0.3)$ & & $v_0 \sim \Gamma(220,0.5)$ \\
& &  $v_1\sim \operatorname{Exp}(1 \diagup 5)$\\
& &  $\sigma_1\sim \operatorname{Exp}(1 \diagup 5)$ \\
&   & \\
$D\sim \operatorname{U}(2.5,4.5)$ & $D\sim \operatorname{U}(2.5,4.5)$ & $D\sim \operatorname{U}(2.5,4.5)$ \\
$S\sim \operatorname{U}(0.25,1.25)$ & $S\sim \operatorname{U}(0.02,0.05)$ & $S\sim \operatorname{U}(0.01,0.03)$ \\
$\sigma_y\sim \operatorname{Exp}(1 \diagup 0.1)$ & $\sigma_y\sim \operatorname{Exp}(1 \diagup 0.1)$ & $\sigma_y\sim \operatorname{Exp}(1 \diagup 0.1)$ \\
\hline
\end{tabular}
\end{table}

We present results from two datasets simulated from SM91: one in which data are from an unforced
version, denoted SM91-u, in which parameters $\gamma_P=\gamma_C=\gamma_E=0$ so that $F=0$, and a
forced version, SM91-f, for which these parameters and $F$ are non-zero.  The parameter values used
were: $p=0.8$, $q=1.6$, $r=0.6$, $s=1.4$, $v=0.3$, $\sigma_1=0.2$, $\sigma_2=0.3$, $\sigma_3=0.3$,
$D=3.8$, $S=0.8$, $\sigma_y=0.1$, and additionally for SM91-f $\gamma_P=0.3$, $\gamma_C=0.1$, 
$\gamma_E=0.4$, which are comparable with those estimated from real data.
We simulate observations every $3$ kyr over the past $780$ kyr to give 261 observations in each dataset,
 comparable to a  low resolution sediment core.  From these datasets we calculated 
the model evidence and posteriors for each of five models: the forced and unforced versions of SM91 
and T06, and the forced model PP12. We do not consider an unforced PP12 model because the deglaciation-glaciation
transition depends only on the astronomical forcing (whereas SM91 and T06 both oscillate in the
absence of any external forcing). The models contain between 10 and 16 parameters. We then test the
ability of  our inference algorithms to 1) discriminate between the five models by  estimating  the
Bayes factors; and 2) recover the parameters used to generate the data.
The priors used for each model are given in Table~\ref{Tab:Priors}. 

\begin{table}
\caption{\label{Tab:EvidenceSS}Log Bayes factors for comparing five different models on the two simulated datasets. SM91-u is data generated from an unforced version of SM91, whereas SM91-f is generated from an astronomically forced version of SM91.}
\centering
\begin{tabular}{l l c c}
\hline
\multicolumn{2}{c}{Model} & \multicolumn{2}{c}{Dataset} \\
%\cline{3-6}
& & SM91-u & SM91-f\\
\hline
SM91 & Forced & $-1.6$ & $0$\\
 & Unforced & $0$ & $-22.5$ \\
T06 & Forced & $-9.8$ & $-10.4$   \\
 & Unforced & $-8.0$ & $-26.4$ \\ 
PP12 & Forced & $-20.7$ & $-21.4$ \\
\hline
\end{tabular}
\end{table}

The estimated $\log_{10}$ Bayes factors ($\log_{10}$ BF) 
 are given in Table~\ref{Tab:EvidenceSS}. 
A common interpretation suggests that $\log_{10} B_{12}\approx 1$ is strong evidence in favour  of  model $\mathcal{M}_1$ over model $\mathcal{M}_2$, and that  $\log_{10} B_{12} \approx 2$ is  very strong evidence that $\mathcal{M}_1$  is superior \citep{Kass1995}. Conversely, a negative score indicates the same strength of evidence but in the other direction (for $\mathcal{M}_2$ over $\mathcal{M}_1$).
 In each column, the $\log_{10}$ BF is with respect to the true generative model, so that positive values indicate support for that model over the true model, and negative values indicate support for the true model. 
Because the $\log_{10}$ BF is just the difference between the log evidences, we can reconstruct the evidences by noting that the $\log_{10}$ evidence ($\log_{10}\pi(y_{1:M}|\mathcal{M})$) is 28.5 for the unforced version of SM91 on the SM91-u dataset, and 40.9 for the forced SM91 model on the SM91-f dataset. 

For both simulated datasets, we find a strong preference for the correct model.  When applied to SM91-f,
the correct model (the forced SM91 model) is overwhelmingly favoured. The $\log_{10}$ BF to the next most
supported model (the forced T06 model) is estimated to be 10.4,  indicating decisive evidence in favour of
the true model.  It is interesting to note that if we remove the forced SM91 model from the analysis, we
find decisive evidence in favour of the forced T06 model over any of the other unforced models (a $\log_{10}$ BF
of at least 11), showing that the astronomical forcing has explanatory power even in the wrong model (to find other BFs, note that $\log B_{ij} = \log B_{i0}-\log B_{j0}$).
This is not particularly surprising, because in both models the astronomical forcing acts as a
synchronisation agent, controlling the timing of terminations, and has a strong effect on the likelihood.
This is a reassuring finding: it suggests that palaeoclimate scientists can implicitly rely upon 
this effect when 
arguing for the importance of the astronomical forcing, as it allows us to infer its importance even when
using an incorrect model (for we surely are).

When applied to SM91-u the log BF again correctly identifies the correct generative model, although the support for the unforced and forced SM91 models is now much closer (with a $\log_{10}$ BF of 1.6 in favour of the unforced model). In cases where the forcing does not add any explanatory power this is an expected result, as the unforced version of SM91 is nested within the forced version, and can be recovered by setting $\gamma_P = \gamma_C = \gamma_E = 0$. This effect is also noticeable when comparing the forced and unforced T06 models, with the unforced version being preferred with a $\log_{10}$ BF of 1.8. 

These experiments clearly show  that there is  sufficient information in the data to easily detect the
correct parametric form of the model in each case. Note that  care needs to be taken when using Monte Carlo estimates of the model evidences, as the Monte Carlo error can be considerable.
Experimentation suggests \citep{Carson2014} that the model evidence estimates have a variability of approximately an order of magnitude, and hence the $\log_{10}$ Bayes factors should be viewed as having variability of approximately plus or minus 2 on the $\log_{10}$ scale. Hence, taken together with the suggested interpretation of BFs in \citet{Kass1995}, this suggests the conservative rule-of-thumb that the $\log_{10}$ BF should be at least 3 to constitute strong evidence for one model over another.
Note that our conclusions from the simulation study are mostly unaffected by this noise, aside from further confirming that the difference between the forced and unforced version of the same parametric model is small on the unforced dataset.
The magnitude of the estimation error can be decreased by using more particles in the SMC$^2$ algorithm, but this will require very long computational runs. Using $N_x=N_\theta=1000$ takes 3-4 days on a standard desktop depending on the model. However, SMC algorithms are well suited to run in parallel, and we were able to obtain a ${\sim}25\times$ speed-up on a Tesla K20 GPU. 
 
The marginal posterior distributions for the parameters for the forced SM91 model applied to the SM91-f dataset are shown in Figure~\ref{Fig:PosteriorSS}.
We are able to recover the parameters used to generate the data, with the true values lying in regions of high posterior probability.
The posteriors for $q$ and $\sigma_3$ do not deviate much from the prior, suggesting that a wide range of values explain the data equally well.
Further simulation studies and details are available in \citet{Carson2014}.

\subsection{ODP677}
\label{SS:677}

We now analyse data from the ocean drilling programme (ODP). The estimated $\log_{10}$ BFs for each model are given in Table~\ref{Tab:EvidenceODP}. ODP677-u refers to an age  model derived by \citet{Huybers2007} using a depth derived model, whereas ODP677-f is an astronomically tuned age model described in \citet{Lisiecki2005}.
The BFs are given in comparison to the best model for each dataset.  
For ODP677-u, the unforced T06 model is best supported, but the estimated BF for the unforced T06 model compared to the unforced SM91 model is within our Monte Carlo bounds, and so it is not possible to confidently assert that T06 is superior to SM91 for explaining these data. There is reasonable to strong evidence (given our Monte Carlo uncertainty) that the unforced models are preferred to the forced models, i.e., there is reasonable evidence that we do not need an astronomically forced model to explain the data. This resembles the simulation study results for SM91-u, where the forced models are penalised for containing extra parameters with little explanatory power. 
That the unforced model is preferred may be surprising compared to earlier works based on similar records \citep{Raymo97, Huybers2011}; this  is discussed further in the conclusions.

\begin{table}
\caption{\label{Tab:EvidenceODP}Log Bayes factors for comparing five different models on ODP677. Values of the log evidence can be reconstructed using $\log_{10} \pi(y_{1:M}|\mathcal{M})= 28.2$ for the unforced version of T06 on the ODP677-u dataset, and $\log_{10} \pi(y_{1:M}|\mathcal{M})=33.7$ for  PP12  on the ODP677-f dataset. }
\centering
\begin{tabular}{ l l  c c  }
\hline
\multicolumn{2}{c}{Model} & \multicolumn{2}{c}{Dataset} \\
%\cline{3-6}
& & ODP677-u & ODP677-f\\
\hline
SM91 & Forced &  $-3.2$ & $-5.8$ \\
 & Unforced &  $-1.7$ & $-15.5$ \\
T06 & Forced & $-2.9$ & $ -4.0$  \\
 & Unforced &  $0$ & $-12.2$  \\ 
PP12 & Forced & $-3.8$ & $0$  \\
\hline
\end{tabular}
\end{table}

When we analyse ODP677-f, the astronomically tuned data, the results are reversed. We now find that the PP12 model is strongly indicated by the data, and that the three forced models are all decisively preferred to the two unforced models, i.e., we find overwhelming evidence using these data that astronomical forcing is necessary to explain the data. 
The orbital tuning of ODP677-f is the most likely explanation for this.
In SM91 and T06, the astronomical forcing acts as a pacemaker, controlling the timing of glacial inceptions and terminations,
while in PP12 the astronomical forcing dictates the transition from the glaciated state to the deglaciated state.
As such, we might expect the output of PP12 to be more strongly correlated to the astronomical forcing, as found for  ODP677-f.
Forced SM91 and forced T06 are both more supported than the unforced versions, with large Bayes factors, but
Monte Carlo error makes it difficult to judge whether T06 is more supported by the data than SM91.

This result is our second key finding. Namely, that inference about the best model is strongly affected by the age model used to date the data. It is vital that modelling assumptions in the dating methods should be understood when performing inference on palaeoclimate data. Given that the two chronologies, ODP677-f and ODP677-u, are considered consistent  once we account for dating uncertainties, we suggest that this formally demonstrates that the approach of first dating the data, and then carrying out down-stream analyses given this dating (ignoring the uncertainty) may  undermine any subsequent inference about the dynamic mechanisms at play.

\begin{figure}
\centering
\includegraphics[width=\textwidth]{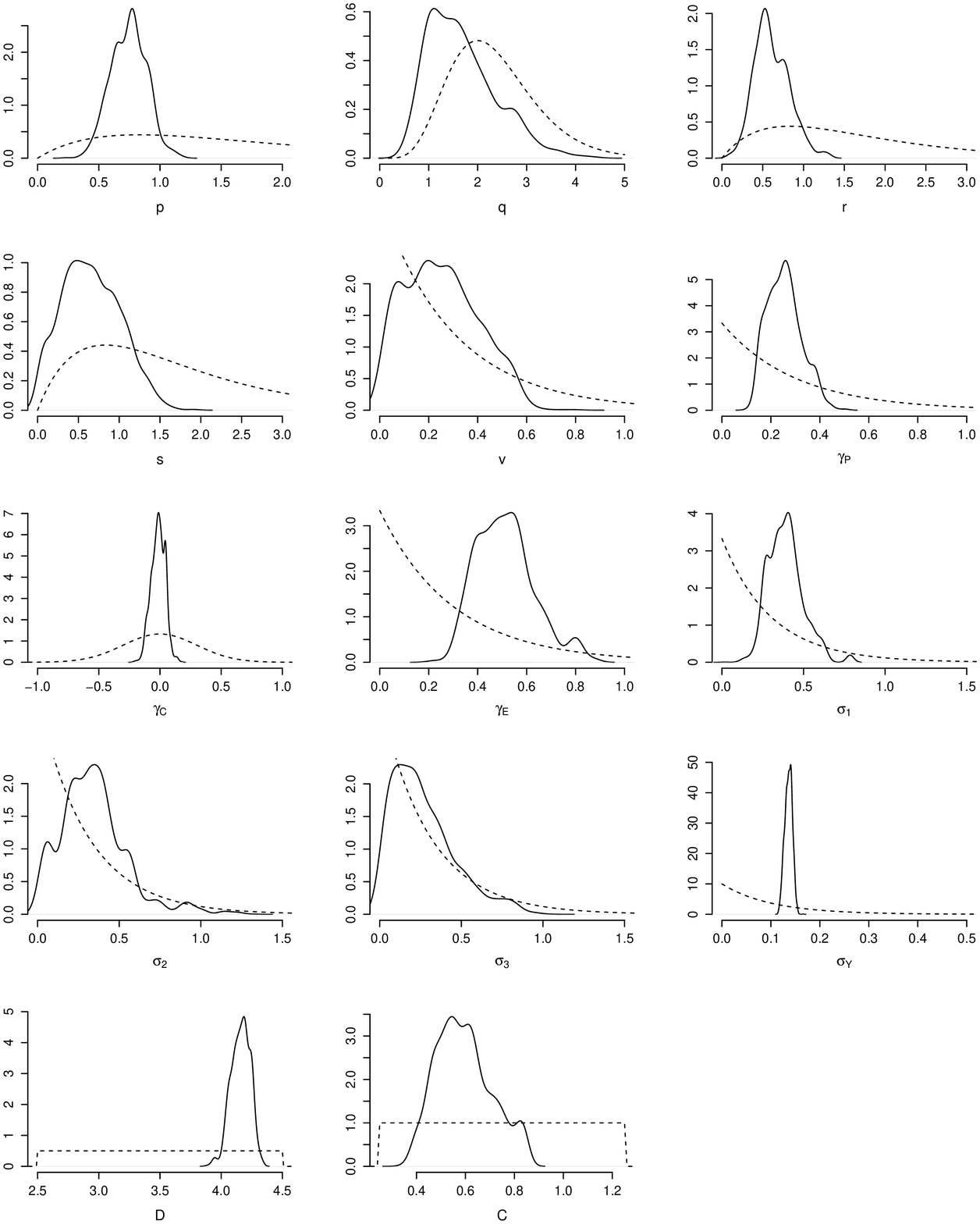}
\caption{
Marginal posterior distributions for the fully forced SM91 model on ODP677-f. Dashed lines represent the prior distributions.}
\label{Fig:PosteriorODP}
\end{figure}

The marginal posterior distributions of the parameters in the SM91 model when fit to the ODP677-f data are shown in Figure~\ref{Fig:PosteriorODP}.
The astronomical forcing scaling parameters $\gamma_P$ and $\gamma_E$ have very small posterior probabilities at 0, suggesting that both precession and obliquity are important.

\begin{figure}
\centering
\includegraphics[width=0.45\textwidth]{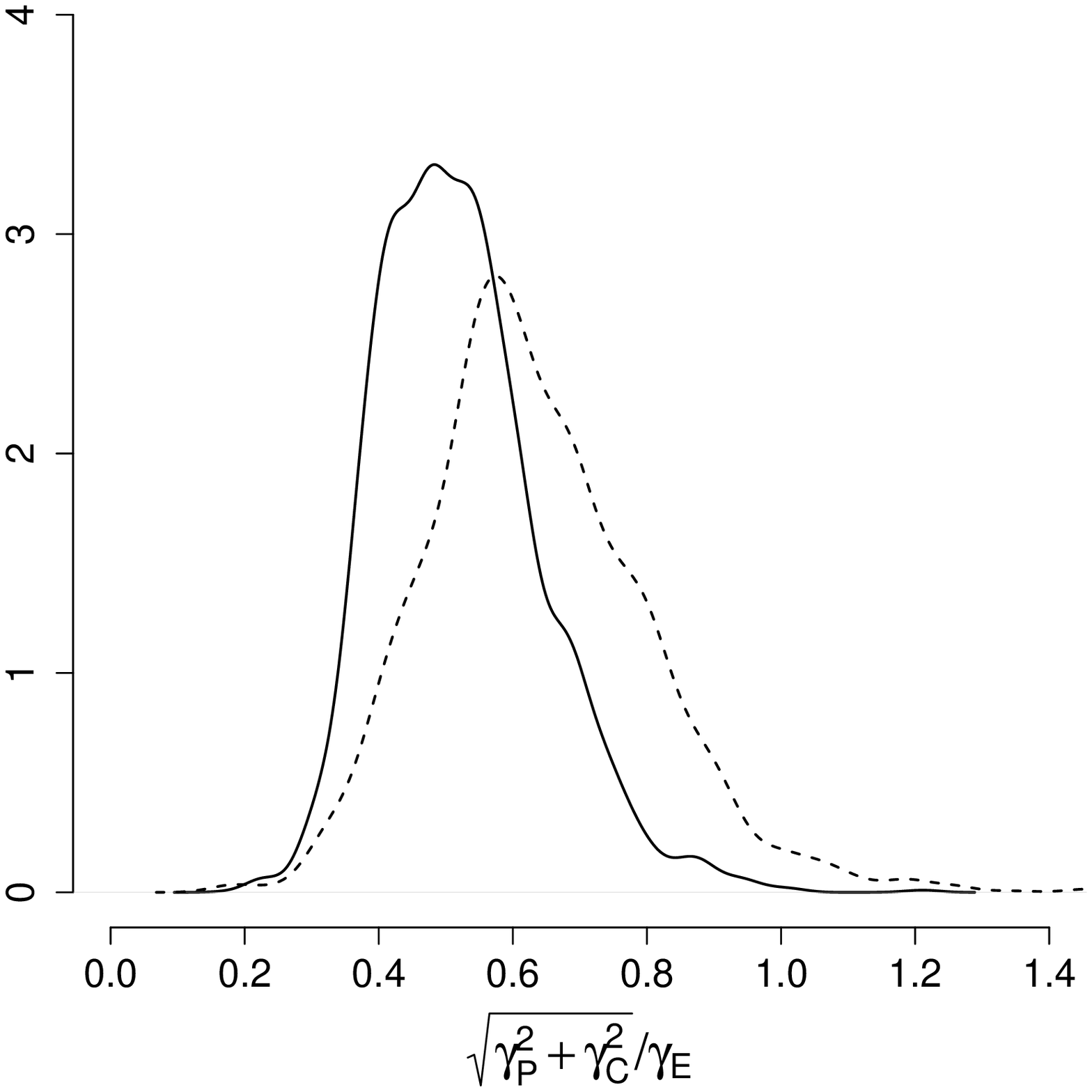}
\includegraphics[width=0.45\textwidth]{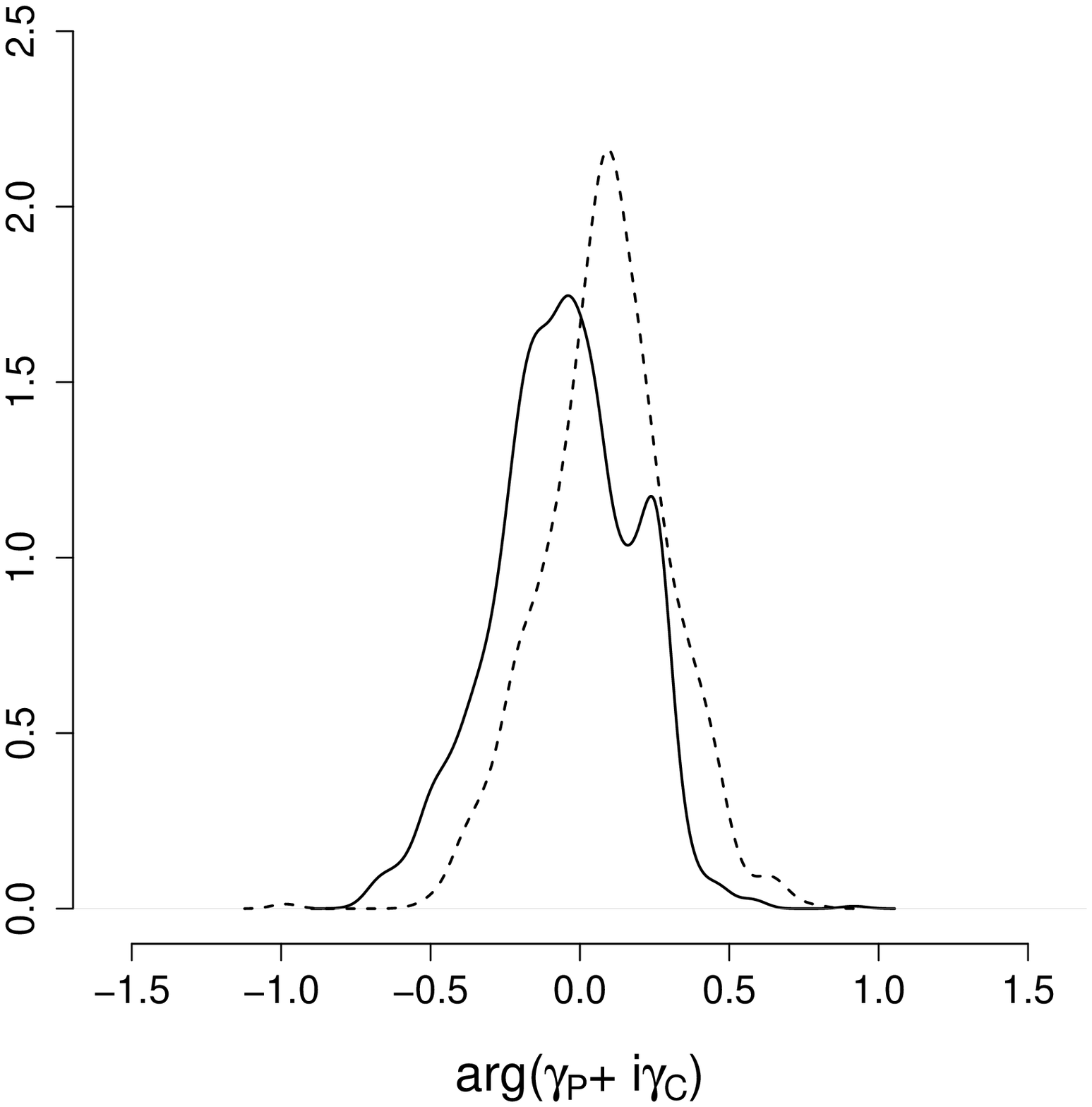}
\caption{
Posterior density of the relative contribution between precession and obliquity in the astronomical forcing (left), and the phase of the precession (right) for the SM91 model (solid line), and T06 model (dashed line).}
\label{Fig:AFRatio}
\end{figure}

Figure~\ref{Fig:AFRatio} provides the density of the ratios $\frac{\sqrt{\gamma_P^2  + \gamma_C^2}}{\gamma_E}$, which measures the relative weights of precession and obliquity in the forcing. 
PP12 is omitted as the truncation of the forcing makes  the parameters incomparable.
Results are consistent across models, with a ratio lower than one, suggesting that the control of obliquity dominates. Translated in terms of palaeoclimate dynamics, this means that
ice age dynamics are controlled by insolation integrated over a season length, rather that just the maximum insolation over the year.
Figure~\ref{Fig:AFRatio} also shows the argument of the complex number   $\gamma_p + i \gamma_c$.  Zero means that phase of the precession forcing 
matches that of the June solstice insolation. A phase of $\pi/2$ would mean that the system is  controlled by March insolation, while
 $-\pi/2$ would point to September insolation. All densities are broadly centred on zero, suggesting a summer insolation control in the Northern Hemisphere  (a winter control in the Southern Hemisphere would be equally consistent, but physically less plausible). The nominal uncertainty of approximately 0.8 radians translates into an uncertainty of about 2 months in calendar time. Physically, it is reasonable to assume that the driving effect changes as ice sheets grow and melt, and that these changes contribute to the variance of the density curves. We acknowledge that dating assumptions will also presumably be crucial in determining this quantity.
  
\section{Conclusions}
\label{S:Conclusion}

We have two key conclusions.  The first is that Monte Carlo technology and computer power are now both
sufficiently advanced that with work, it is possible to fully solve the Bayesian model selection problem
for a wide class of phenomenological models of the glacial-interglacial cycle. It came as a surprise to
us, that even relatively short time series of observations contain sufficient information to discriminate
between many of the models. A priori, we had expected to find that there was simply insufficient
information in the data, given the level of noise, to solve simultaneously the filtering and 
calibration problems, and  still distinguish between the models. That we are able to do this,
contrasts strongly with the  viewpoint set out in \citet{Roe1999}:

\begin{quote}
Most simple models of the [...] glacial cycles have at least four degrees of freedom [parameters], and some have as many as twelve. Unsurprisingly [...this is] insufficient to distinguish between the skill of the various models.
\end{quote}

\noindent In 1999 this viewpoint may have been true. We lacked both the computer power and the algorithmic knowledge to do Bayesian inference for the  parameters, never mind  estimating the Bayes factors. However, recent developments in Monte Carlo methodology, and the massive increase in computing power (including the utilisation of GPUs), means that the calculations are now possible. Using only 261 observations, we are able to learn up to 16 parameters,  state trajectories  containing $261\times 3$ values, and  calculate the marginal evidence. Moreover, these evidences are sufficiently different (and able to be estimated with sufficient accuracy) that we can confidently discriminate between the ability of the models to explain the data.
Of course, that one dynamical system is more supported by the data than another does not necessarily imply that the \emph{physical} interpretation of that model is valid. At this level of conceptual modelling, different physical interpretations may produce similar equations. This point has been made before \citep{Tziperman2006} 
and we add here that the stochastic differential equations emerge as a combination of judgements on physical processes \emph{and} model discrepancy, embedded in the stochastic parameterisations. 
On the other hand, the Bayesian formalism for choosing between models offers a natural starting point for developing a physical interpretation, and knowledge  of physical constraints can be incorporated within the parameter prior distributions. Physical disambiguation will also arise as  the complexity of the model is increased, and as more diverse datasets are used in the inference process.   

Our second conclusion concerns the need to avoid ``theory-laden'' data. The results from analysing the  ODP677 data, show that the age model used to date the core become critical when the data are subsequently used to make scientific judgements. The astronomically-tuned age model gives support for a model in which ice ages are \textit{driven} by the astronomical forcing (that is, without an underlying autonomous limit cycle), while the age model which was not tuned on astronomical forcing favours models explaining ice ages as an autonomous limit cycle. These are two qualitatively different explanations of ice ages. Admittedly,  fifty years of climate research  have established beyond doubt  that the astronomical forcing affects the climate system enough to interfere with ice ages dynamics: the rejection of astronomical forcing here must presumably be explained by errors in the ODP677-u time scale. On the other hand, we observed that both time scales were compatible with uncertainties provided by the respective authors. 
This suggests that analysing the data in stages, cutting feedbacks between uncertainties, does not only affect the conclusion about the role of the astronomical forcing,  it also affects inferences about the internal system  dynamics. We believe that this is the first time the effect of the age model on subsequent analyses has been so clearly demonstrated.
Instead of first dating the core, and then using those dates (with or without uncertainties), we  need to jointly estimate the age model at the same time as testing further hypotheses, accounting for all the joint uncertainties. 

The experiments included in this paper can be extended in several ways.
Firstly, we considered only a handful of models, and both the number and complexity of models can be increased. 
With the approach described here, extra models can be included by running the SMC$^2$ algorithm for each model.
This has the benefit that the entire experiment does not need to be redesigned/repeated for different combinations of models.
Different astronomical forcings can also be considered.
For example, the astronomical forcing terms are often tested independently.
This can easily be achieved by setting undesired astronomical scaling terms to 0 in our forced models.
Making the forcing term state dependent, so that an increase in sea-ice increases albedo, which in turn alters the influence of variation in insolation, is also a possibility. 
%\mcr{it is not so much a problem of sea-ice but, for example, the ablation area grows non-linearly with insolation; there are references for this but we can see this later.}

Finally, we do not need to limit ourselves to a single dataset.
The observation model can be extended to compare the state of the system to multiple cores.
Likewise, multivariate observations could be used; SM91 models both ice volume and CO$_2$ concentration, and records exist for both of these quantities.

Overall, we hope that this work acts as a proof of concept. Careful statistical analysis combining data and models can lead to insights in palaeoclimate science.

\bibliographystyle{plainnat}
\addcontentsline{toc}{chapter}{References}
\bibliography{FullReferences}

\end{document}